# Peculiarity of chaotic and regular dynamics of waves

**V.A. Buts, I.K. Kovalchuk, D.V. Tarasov, A.P. Tolstoluzhsky**
National Science Center "Kharkiv Institute of Physics and Technology", National Academy of Sciences of the Ukraine,
Akademicheskaya Str. 1, Kharkiv, 61108 Ukraine
Email: vbuts@kipt.kharkov.ua

**Abstract**

It is shown, that at weakly nonlinear interaction of waves are possible as modes with chaotic dynamics, and with increasing degree of coherence. Conditions are found at which they arise. One of the types of such interaction is decays. The important features of such processes in plasma are modes with cascades. They arise in that case when the high-frequency wave which has appeared as a result of decay can take part in new decay. Chaotic dynamics of decays can be used for formation of spectra of excited oscillations. For realization of such possibility the dispersive properties of a cylindrical wave guide partially filled by magnetoactive plasma are investigated. It is shown, that such electrodynamic structure is convenient for realization as regular and chaotic modes.

**Keywords:** regular; chaotic dynamic; entropy; dispersion properties; magnetoactive plasma; decay

## 1. Introduction

In plasma it is possible to define two basic physical processes - interaction of a wave-particle and a wave-wave type. In this paper we present the results of the regular and chaotic dynamics of three-wave interaction, a cascade of such interactions and also the use of these processes for formation of spectrums of excited oscillations. In addition we considered the opportunity of increasing the degree of coherence at three-wave interactions.

By present time it is enough investigated the regular dynamics of three-wave interaction. Feebly explored is the process which proceeds simultaneously both in space and in time (it is described by the nonlinear partial differential equations). The process with participation of completely not correlated waves is in also enough investigated in details. In this case the use of approach of chaotic phases allows to reduce the problem to studying the dynamics of intensity. Thus there always has been the question on opportunities when the use of this approach is possible.

In papers ([1-6]) it has been shown that the process of the regular nonlinear interaction can become chaotic if the amplitude of the breaking up wave will exceed some threshold. The criterion of occurrence of chaotic dynamics which allows to specify the requirements at which the use of approach of chaotic phases is possible, has been obtained.

It is known that in many cases, in particular, at realization of the beat-wave scheme for acceleration of the charged particles, the laser incident wave on plasma breaks up on another high-frequency wave and a low-frequency Langmuir wave. The new high-frequency wave excited in such scheme can break up in turn on a new high-frequency Langmuir wave. Such process can iterate. There is a stage of three-wave decays. For the first time the regular dynamics of such infinite decays has been investigated in papers ([7-8]). In real situations the process of decay contains the finite number of interacting waves. As it will be shown below (section 2), the presence of the finite number of decays leads to qualitative change of dynamics, in particular, to chaotic dynamics.

The interesting feature of three-wave interaction of waves is that during such interaction the degree of coherence of a high-frequency wave can be increased (section 1).

The processes of decay with chaotic dynamics can be used for broadenings the spectrum of interacting waves. Such broadening allows to create the source of radiation with a wide spectrum. To define of a real opportunity of realization of such broadening in section 3 the variance of property of a cylindrical wave guide partially filled with rare magnetoactive plasma was investigated. It is shown, that near the upper hybrid resonance there is dense enough spectrum of natural waves which can be used with success for realization the regimes with stochastic decay.

## 2. The increase of degree of coherence at three-wave interaction

Above we said that the weak nonlinear wave-wave interaction can be both regular and chaotic. Occurrence of chaotic dynamics of processes of three-wave interaction was revealed and studied in papers [1-6].

In the present section we shall show, that at three-wave interaction can take place and processes, at which the degree of coherence some of the waves participating in interaction, can essentially will increase.

In order to prove this possibility we will consider the process of decay of high-frequency wave on high-frequency and low-frequency ones. The regular process of such decay is studied rather well [9,10]. We will consider that at the initial moment the non-monochromatic high-frequency wave propagates in non-linear medium. The spectrum width of this wave is equal $\Delta\omega$. Besides, to simplify the further analysis, we will consider that the source of irregularity of this wave is random diffusion of a phase. For such model it is important that the equation for a random phase can be written in the following way:
$$\dot{\varphi} = \omega_1 + \xi(t), \quad (1)$$
Where $\xi(t)$ - delta-correlated function with a zero average:
$$\langle \xi \rangle = 0, \quad \langle \xi(t) \cdot \xi(t') \rangle = \delta\omega \cdot \delta(t - t'). \quad (2)$$

The function $\omega_1 + \xi(t)$ is the instantaneous signal frequency. The diffusion coefficient $\delta\omega$ is spectrum width of a signal. Except high-frequency wave, in non-linear medium there are small coherent high-frequency perturbations, and also small low-frequency perturbations. Besides we will consider, that the amplitude of the main signal is large enough, so the increment of decay instability is larger than the spectrum width of this signal ($\Gamma > \delta\omega$). In this case the equations describing the dynamics of complex amplitudes of the interacting waves can be presented in the following way:
$$\dot{A}_1 = -A_2 \cdot A_3, \quad \dot{A}_2 = A_1 \cdot A_3^*, \quad \dot{A}_3 = A_1 \cdot A_2^*, \quad (3)$$

While obtaining these equations the laws of conservation of energy and impulse have been taken into account:
$$\sum_{i=1}^{3} \omega_i N_i = const \quad \sum_{i=1}^{3} \vec{k}_i N_i = const. \quad (4)$$

Besides, the normalization has been used: $|A_k|^2 = N_k$ - number of quanta in $k$ - wave; $\dot{A} = dA/d\tau$, $\tau = V \cdot t$, $V$ - a matrix element of wave coupling. The equations (3) have the following integrals:
$$|A_1|^2 + |A_2|^2 = const, \quad |A_1|^2 + |A_3|^2 = const,$$
$$|A_2|^2 - |A_3|^2 = const \quad (5)$$

At the initial stage of decay process it is possible to consider that the amplitude of a breaking up wave does not change. In this case it is enough to analyze the last two equations of the system (3). Thus, it is convenient to introduce the new function $B = A_1 \cdot A_3^*$. Then the last two equations of the system (3) can be rewritten in the following way:
$$\dot{A}_2 = B, \quad \dot{B} = \xi \cdot B + |A_1|^2 \cdot A_2. \quad (6)$$

Using the formula of Furutsu-Novikov [11], and also that ($\delta B(\xi)/\delta\xi(t) = \delta\omega \cdot B(t)$ where $\delta B(\xi)/\delta\xi(t)$ - variational derivative), from system (6) it is possible to obtain following equations for the first moments (average values):
$$\langle \dot{A}_2 \rangle = \langle B \rangle, \langle \dot{B} \rangle = |A_1|^2 \cdot \langle A_2 \rangle + \delta\omega/2 \cdot \langle B \rangle. \quad (7)$$

From (7) it can be seen that the process of decay for average values takes place also (practically with the same increment $\Gamma = \sqrt{|A_1|^2 + (\delta\omega)^2/16} + \delta\omega/16 \approx |A_1|$) as well as at decay of the regular wave ($|A_1|^2 \gg \delta\omega$). The fact of the regular dynamics of decay for average values in our case allows to conclude that the fluctuations of the phase of a breaking up wave $A_1$ are compensated by the fluctuations of the phase of a low-frequency wave $A_3$. Really average value of function the $B(\tau)$ has the same dynamics as the regular function. As a result, the fluctuations of complex amplitude $A_2$ and function $B(\tau)$ are small. For further analysis it is necessary to consider the non-linear dynamics. For this purpose the set of equations (3) it is convenient to rewrite in the following:
$$\frac{d(|A_1|^2)}{d\tau} = -[A_2 B^* + A_2^* B], \quad \dot{A}_2 = B, \quad (8)$$
$$\dot{B} = A_2 [2|A_1|^2 - |A_1(0)|^2]$$

Taking into account the obtained above result that fluctuations of amplitude $A_2$ and function $B(\tau)$ are small $A_2$, it is easy to average the set of equations (8). It is seen that in this case these average equations will coincide with the equations (8). In turn, the equations (8) do not differ from the equations which describe the regular dynamics of the decay process. Thus, as it is well known, during the time which is order of a return increment ($T \sim \Gamma^{-1}$) almost all the energy of a breaking up wave transfers into the energy of high-frequency coherent component. The negligible part of the energy, according to Manley-Rowe's relation, transfers into a low-frequency wave ($\delta E \sim (\Omega/\omega_1) E \ll E$). Let's notice that according to integrals (5), the number of quanta of a low-frequency wave is equal to the number of quanta of high-frequency waves.

Thus, we have shown that, despite that the phase of a breaking up wave undergoes random fluctuations, this wave can transmit almost all the energy to monochromatic coherent high-frequency wave, and a part of the energy will transfer to a low-frequency wave. Now, if we interrupt non-linear wave coupling after the full pumping of the energy from an initial wave into a new high-frequency one and in the field of a low-frequency wave, then the process will appear as transformation of the energy of an incoherent wave into the energy of coherent wave. It can seem that such process proceeds with violation of the second law of thermodynamics. However, as it is shown in paper [12], such processes proceed with entropy growth in the full conformity with the second law of thermodynamics. The explanation is such: in this case all entropy from high-frequency wave transforms into entropy of a low-frequency wave.

Really, it is easy to show that the full entropy (the entropy of high-frequency waves plus entropy of low-frequency wave) will be equal or even exceed the entropy of an initial wave. In order to prove this fact we will write down the expression for the entropy which the breaking up wave had during the initial moment of time and we will define the entropy of low-frequency waves. The entropy for each of interacting waves (bosons) can be written down (see, for example [13]) in the following form:

$$S_{t,l} = G_{t,l}\left[(n_{t,l}+1)\cdot\ln(n_{t,l}+1) - n_{t,l}\cdot\ln(n_{t,l})\right]$$

Where $G_{t,l}$ is the number of quantum states for transversal and longitudinal waves accordingly. This number for transverse waves is equal to:

$$G_t = \Delta g_t \cdot \delta\theta_t = \omega_1^3 \cdot \delta\omega \cdot \delta\theta_t / 8\pi^3 c^3, \qquad (9)$$

Where $\Delta g_t$ - the number of oscillators of the field inside the frequency interval $\omega, \omega+\delta\omega$; $c$ - wave velocity in medium, $\delta\theta_t$ - the spatial angle in space of wave vectors. In this equation $n_j = N_j/G_j$ - the average number of particles in each of quantum states.

For definiteness we will consider the transverse wave decay ($\vec{k}_1$) on transversal ($\vec{k}_2$) and on Langmuir ($\vec{\kappa}$). The scheme of such decay is presented in Fig. 1.

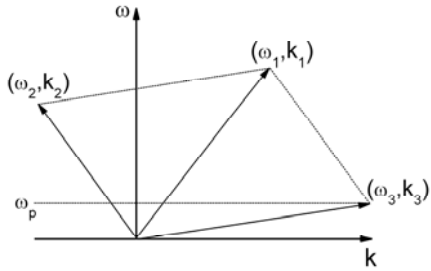

Fig. 1. The dispersion diagram of interacting waves. HF wave (1) decays on new HF wave (2) and LH one (3)

In this Fig. it is seen that between wave vectors of interacting waves the such relation (a synchronism requirement) should be carried out $\vec{k}_1 = \vec{k}_2 + \vec{\kappa}$. From this relation it is possible to find at once the relationship between the phase velocity of high-frequency waves $c$ and phase velocity of a low-frequency wave: $v \approx c \cdot \Omega/2\omega$ and also $\delta\theta_t \sim 4\cdot\delta\theta_l$ (look Fig. 2, $|\vec{\kappa}| \sim 2|\vec{k}_1|$). We will consider that the high-frequency wave which is excited due to the decay is monochromatic, i.e. its entropy is equal to zero. Fluctuations of the phase of a breaking up wave have transferred into fluctuation of the phases of low-frequency waves. Therefore the characteristic width of the spectrum of low-frequency waves is equal to the width of the spectrum of breaking up high-frequency wave. Considering these facts, it is easy to make the estimation: $G_l = \Delta g_l \cdot \delta\theta_l = \Omega^3 \cdot \delta\omega \cdot \delta\theta_l / 8\pi^3 v^3 \sim 2G_t$. Considering also that fact that the number of low-frequency quanta is equal to the number of high-frequency ($N_t = N_l$), we find out that $n_l = n_t/2$. Substituting the expressions for the number of quantum states and for the density of these states it is easy to see that the entropy of low-frequency waves exceeds the entropy of breaking up high-frequency wave $(S_l > S_t)$.

It's necessary to note that this effect does not depend on the density of states $n_{t,l}$. Really in Fig. the relation of the quantities of entropy of longitudinal waves to the quantity of entropy of transverse waves is presented.

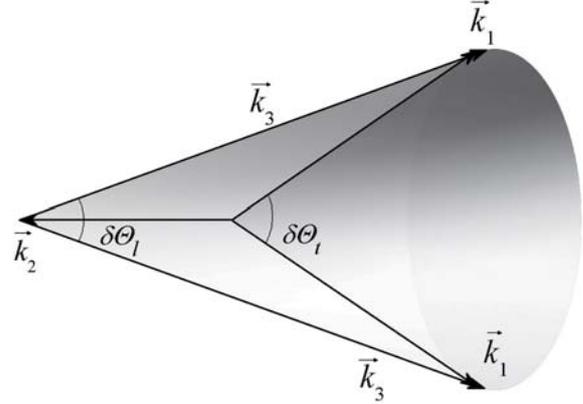

Fig. 2. To definition of entropy of HF and LF waves. HF wave with a coefficient 1 breaks up on backward HF wave with coefficient 2 and LF one with coefficient 3 (see Fig. 1). Opposite to Fig.. 1 the angular characteristic of interacting waves are presented.

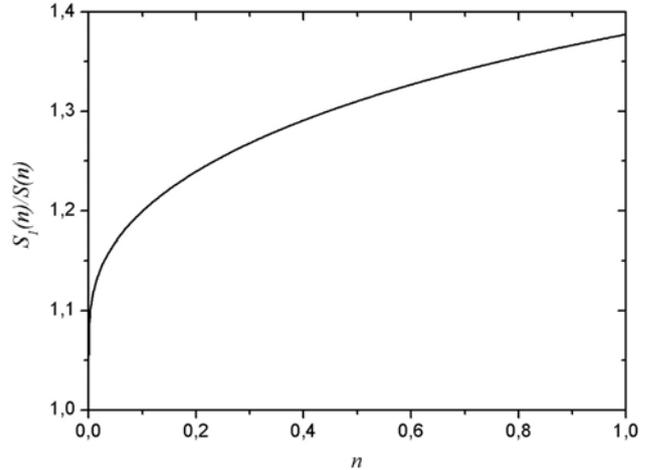

Fig. 3. The relation of value of an entropy of longitudinal waves to the value of an entropy of transverse waves. $n$ - the average of quantums in each of the states. As it is seen in the Fig. – the value $S_l$ always large then $S_t$.

## 3. The dynamics of cascades in consideration of processes of merging.

In the theory of weakly-nonlinear interaction of waves the model which is widely used at the analysis of the processes arising at influence of powerful laser radiation on plasma is known. In it the unlimited number of interacting waves can be considered. We shall consider, that the number of high-frequency waves can be unrestrictedly large, and interaction between them is carried out through one low-frequency wave. Thus will be considered not only processes of a birth of "red" satellites (decay), but also the processes leading to birth of "blue" satellites (merge of waves). Attenuation we shall neglect.

The mathematical model of such processes, apparently, for the first time has been obtained in works [7-8] and analyzed by many authors:

$$i\frac{da_n}{dt} = ba_{n-1}e^{i\delta t} + b^* a_{n+1}e^{-i\delta t} \quad (10a)$$

$$i\frac{db}{dt} = \sum_{n=-N_1}^{N_2} a_{n-1}^* a_n e^{-i\delta t} \quad (10б)$$

Here $\delta = \omega_n - \omega_{n-1} - \Omega$ detuning, $\omega_n$ – frequency of HF wave, $\Omega$ – frequency of LF wave.

In real systems the number of high-frequency interacting waves though is large, but finite. Therefore, we shall suppose that amplitudes of waves $a_n = 0$, if $n$ lies outside the range of values which is determined by condition - $N_1 \leq n \leq N_2$ ($N_1$ - number of "red" satellites, $N_2$ - number of "blue" satellites). The set of equations (10) accurate to terms $\sim O(\Omega/\omega_0, \kappa/k)$ have integrals of motion:

$$\sum_{n=-N_1}^{N_2} \omega_n |a_n|^2 + \Omega |b|^2 = \varepsilon \quad (11)$$

$$\sum |a_n|^2 = I. \quad (12)$$

The relation (2) represents a full energy of interacting waves, and the relation (3) can be interpreted as number of HF waves. Besides, it is possible to show that

$$\frac{d}{dt}\sum_{n=-N_1}^{N_2} a_{n-1}^* a_n = (|a_{-N_1}|^2 - |a_{N_2}|^2). \quad (13)$$

And in case, when the number of "red" and "blue" satellites is equal ($N_1 = N_2$) take place one more integral of motion

$$\sum_{n=-N_1}^{N_2} a_{n-1}^* a_n = const. \quad (14)$$

Really, if at initial time it is given HF wave ($a_0^0 = a_0(t=0)$) and small impurity of LF wave and the number of excited satellites is large ($n \to \infty$) it is possible (see [3–4]) to present solution of the equation (1a) as:

$$a_n(t) = a_0^0 i^n e^{in\beta} J_n(|B|),$$
$$B(t) = \int_0^t b(t')\exp(-i\delta t')dt'; \quad \beta = \arg B \quad .(15)$$

One can easily verify this, by direct substitution (15) in (10a) and using a recursive relation $2J'_n = J_{n-1} - J_{n+1}$. From (15) follows, that the process of satellites excitation is symmetric with respect to changing $n \to -n$. Therefore $|a_n| = |a_{-n}|$, that confirms existence of the integral of motion (14).. As can be see from (15), more and more high harmonics $a_n$ become excited during the time due to interaction with LF wave.

In considered conditions the investigated system has the analytical solution. Dynamics of all interacting waves at this is regular. Thus, we have of the unique enough case of completely integrated system having infinite number of degrees of freedom. From equations (10) it is possible to receive the equation for amplitude of low-frequency wave as:

$$\ddot{b}_1 - i\delta\dot{b}_1 + \omega_{N_L}^2 b_1 = 0, \quad (16)$$

were $\omega_{N_L}^2 = |a_{N_1}|^2 - |a_{N_2}|^2$.

Let's analyze the solutions of this equation in the most simply cases. If the number of «red» satellites is equal to number of «blue» satellites ($N_1 = N_2$), then $\omega_{N_L}^2 = 0$. The solution of the equation (16) becomes:

$$b(t) = b_0 + \frac{1}{i\delta}\dot{b}(0)(e^{i\delta t} - 1) \quad (17)$$

From (17) follows, that value of amplitude of LF wave remains constant if, at $t = 0$ it is given HF and LF waves ($b(0) = b_0$, $a_0^0 = a_0(0)$) and $\dot{b}(0) = 0$. If at the initial moment of time are set two HF of a wave (beat-wave case) the amplitude of a wave of LF oscillates with frequency $\delta$. When $\delta \to 0$ the amplitude $b(t)$ increase linearly with time. Dynamics of process of the waves interaction in this case should be most simple, however the dynamics has not of the analytical solution.

### 3.1 Numerical analysis

The analytical results received above are applicable in case $n \to \infty$ and do not give a possibility to investigate the dynamics of finite number of interacting waves. For finite case $n$ the set of equations (1) is solved numerically at various initial conditions for fields and detuning parameters $\delta$. We have investigated: temporal dynamics of waves interaction, power spectra ($S_\omega$) of realizations, their autocorrelation functions ($C_f$) and maximal Lyapunov's exponent of the interacting waves system. There were selected the real initial values of the fields: $Rea_0(t=0)=a_0^0$; $Rea_{-1}(t=0)=a_{-1}^0$; $b(t=0)=b_0$. The imaginary parts of the fields and amplitude of other waves at $t = 0$ are equal zero.

For case of equal number of "red" and "blue" satellites - $N_1 = N_2 = 11$ at fulfillment of the synchronism condition ($\delta = 0$) and in case of different from zero of the initial amplitudes of zero high-frequency and low-frequency waves ($a_0^0 = a_0(t=0)$ and $b_0 = b(t=0)$), the numerical analysis shows, that the amplitude of LF-wave does not change. It remains constant. Dynamics the HF-fields remains enough regular (see Fig. 4). Maximal Lyapunov's exponent (Fig. 4g) decreases with time and we can doing conclusion that in this case process of the decay is regular.

The availability of asymmetry in number of excited HF waves (for example, $N_1 = N_2 - 1$) essentially changes the dynamics of amplitudes as HF and LF.

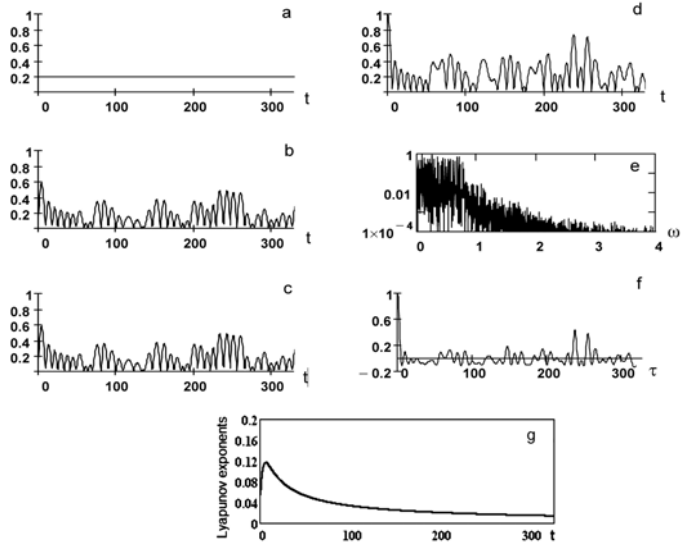

Fig. 4. Dependence of amplitudes of LF of a wave and HF of harmonics on time, spectrum and correlation function at $a_0^0 = 1$, $b_0 = 0.2$, $\delta = 0$. Symmetrical case $N_1 = N_2$. a – LF wave, b – HF wave with $n=1$, c – HF wave with $n=-1$, d - HF wave with $n=0$, e,f – spectrum and correlation function HF wave with $n=0$, g- maximal Lyapunov exponents.

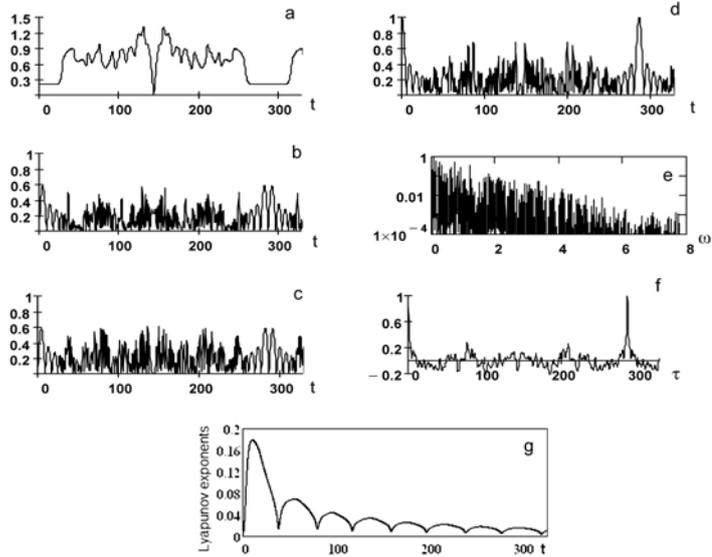

Fig. 5. Dependence of amplitudes of LF of a wave and HF of harmonics on time, spectrum and correlation function at $a_0^0 = 1$, $b_0 = 0.2$, $\delta = 0$. Nonsymmetrical case $N_1 \neq N_2$. a – LF wave, b – HF wave with $n=1$, c – HF wave with $n=-1$, d – HF wave with $n=0$, e,f – spectrum and correlation function HF wave with $n=0$, g – maximal Lyapunov exponents.

In Fig. 5 the results of calculation of dynamics of the fields amplitudes for nonsymmetrical case are shown. The integral (14) is rigorously carried out if number of interacting waves unrestrictedly large big. At the limited quantity of waves it is remain on intervals of time necessary that perturbation has reached last waves participating in interaction. It is interesting to note, dynamics of waves is regularly repeats, including the periodic recovery of integral (14) accompanied by constancy of amplitude of LF-wave takes place. This feature expresses the contents of the Poincare theorem about returns and is similar to the returns observable in a problem of Fermi-Pasta-Ulam.

Spectrum of the HF power has linear character with slow decreasing in the region of high frequencies. Correlation function quickly decreases up to zero and fluctuates near to zero value, and completely comes back to unit value during the moments of time corresponding restoration of integral (14). Maximal Lyapunov's exponents (Fig. 4g) decrease with time and in this case process of the decay is still regular.

When specify at the initial moment of time two high-frequency waves ( $a_0^0$ and $a_1^0$), for realization of the beat-wave type interaction at $\delta = 0$, dynamics of fields is qualitatively similar to the dynamics of fields considered above when at the initial moment of time the amplitude of one high-frequency wave and amplitude of a low-frequency wave (for a case of initial conditions $a_0^0$ and $b_0$) is different from zero.

One more parameter, which essentially influence on the process of waves transformation, is the availability of detuning $\delta \neq 0$. As follows from the previous analytical consideration at small values ($\delta \ll 1$) the dynamics of LF wave insignificantly differs from the case with $\delta = 0$ on times ~ $t < 1/\delta$. At large times the dynamics of non-resonance interaction is investigated numerically. Thus, presence even small detuning ($\delta \ll 1$), leads to irregular dynamics of the waves interaction and occurrence of dynamic chaos, both in a symmetric case ($N_1 = N_2$ and in case of with the broken symmetry ($N_1 \neq N_2$).

On Fig. 6 the results of numerical calculations for asymmetrical case ($N_1 \neq N_2$) are shown with the initial conditions beat-wave $a_0^0=1$ and $a_{-1}^0=0.04$, at detuning parameters $\delta = 0.05$. From these plots it is seen, that the amplitudes of high-frequency oscillations look like the form of irregular oscillations with various intensity, which frequency also varies irregular.

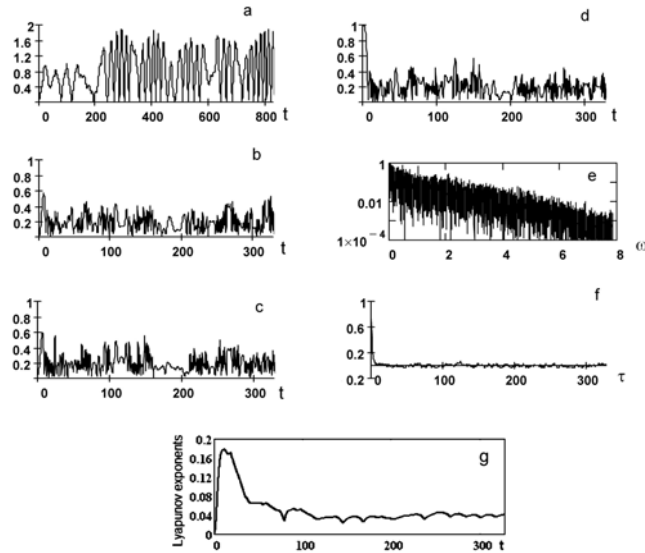

Fig. 6. Dependence of amplitudes of LF of a wave and HF of harmonics on time, spectrum and correlation function at $a_0^0 = 1$, $a_{-1}^0 = 1$, $\delta = 0.05$. Nonsymmetrical case $N_1 \neq N_2$. a – LF wave, b – HF wave with $n$=1, c – HF wave with $n$=-1, d – HF wave with $n$=0, e,f – spectrum and correlation function HF wave with $n$=0, g – maximal Lyapunov exponents

Spectra of these waves acquire the form like «desktop», the correlation functions fast decrease during time and have irregular fluctuations near to this value. Therefore, one can speak about random amplitude and frequency modulation of HF waves. Maximal Lyapunov's exponent with time became practically on constant level (see Fig. 6g) that confirms chaotic character of process. Hence, it is possible to confirm, that at such parameters interaction of waves has irregular character, and dynamics of waves is chaotic.

## 4. Dispersion properties of cylindrical wave guide filled with rare plasma

For many application (for example, for broadband ranging) electromagnetic fields with a broad spectrum are necessary. It is known that for excitation of such broadband noise radiation the special type of generators – the noise generators are created. The theory of such generators, and the technology of their development essentially lags behind the theory and technology of generators of the regular oscillations. It is enough to mention such devices as klystrons, magnetrons, TWT and many others.

These difficulties, first of all, are connected with the fact that in one device it is necessary to create simultaneously requirements for effective transformation of energy of charged particles to the energy of electromagnetic waves, and also requirements for formation of a necessary shape of spectrum of radiation. In overwhelming number of cases these requirements simultaneously cannot be carried out. Thus, it is possible to divide the problem into two parts. First of all, to make highly effective excitation of the regular oscillations, for example, by magnetron and then to generate the necessary shape of a spectrum. The second problem can be solved with a help of regimes with dynamic chaos.

In this case the scheme of excitation of oscillations with a broad noise spectrum can be realized in the following way. Generators of the regular intensive oscillations on an exit create radiation with narrow spectral lines. This radiation arrives on non-linear medium in which modes with dynamic chaos are possible. In this medium the narrow-band radiation of traditional generators can be transformed into the broadband non-regular radiation. However, as numerical estimates show, in usual conditions which are necessary for realization of chaotic modes the intensity of fields becomes very high (see, for example [4,5,6]). So, for example, for the unlimited plasma (for ten-cm band) these intensities of fields are more then 20 kV/sm. It is of interest to find such electrodynamic structures with non-linear elements in which the mode with dynamic chaos would propagate at much smaller intensities of fields.

It is known from the previous investigations that the smaller is the distance of frequency between natural waves of electrodynamic structure, the smaller are the intensities of the fields necessary for transition into regime with dynamic chaos. Therefore our problem is to find out such mediums and structures in which the distance on frequency between natural waves would be minimal. The possible candidate on such structure is the cylindrical metal wave guide which is filled with rare plasma and which is in an external magnetic field. Basically, such electrodynamic structures have been studied before. However the investigations were restricted by studying only slow waves [14]. It is related with the fact that such structures, first of all, were considered to be used for acceleration of charged particles – especially heavy. The dispersion for fast waves was not studied. Latter we will study fast natural waves of such wave guide.

Let's consider an ideal conductive cylindrical wave guide of radius $b$ partially filled with plasma of radius $a$ which axis coincides with a wave guide axis. All system is placed in an external homogeneous magnetic field which is directed along an axis $z$ coinciding with an axis of the cylinder. The dispersion equation for such system is obtained by a standard method as it is stated, for example, in [14,15]. Maxwell equations in a cylindrical coordinate are used for this purpose. The solutions we will present in the form of wave, propagating along axis $z$ : $E, H \sim \exp[i(\omega t - k_z z)]$ where $E$ and $H$ any of the components of the electrical and magnetic fields, $\omega$ – frequency of a wave, $k_z$ – its longitudinal component of a wave vector. Let us consider the axial-symmetric fast waves. The dispersion equation for such waves is:

$$\left(y_1 - \frac{\kappa \Delta_{00}(a)}{\Delta_{10}(a)}\right)\left(y_2 - \frac{\kappa \Delta_{01}(a)}{\Delta_{11}(a)}\right) f_2 -$$
$$-\left(y_1 - \frac{\kappa \Delta_{01}(a)}{\Delta_{11}(a)}\right)\left(y_2 - \frac{\kappa \Delta_{00}(a)}{\Delta_{10}(a)}\right) f_1 = \qquad (18)$$
$$= \frac{\omega_p^2}{\omega^2} \frac{\kappa \Delta_{00}(a)}{\Delta_{10}(a)} \left[ f_1 y_1 - f_2 y_2 + \frac{\kappa \Delta_{01}(a)}{\Delta_{11}(a)}(f_2 - f_1) \right],$$

Where $k = \omega/c$, $y_{1,2} = \left(k_{1,2} J_0(k_{1,2}a)\right)/\left(J_1(k_{1,2}a)\right)$, and $\Delta_{ik}(r) = N_i(\kappa r) J_k(\kappa b) - J_i(\kappa r) N_k(\kappa b)$, $J_i(x)$, $N_i(x)$ - Bessel function and Neumann $i$-th order, accordingly, $k_{1,2}$ - the wave numbers which describe traversal structure of the field in plasma which look as :

$$k_{1,2}^2 = -\left[k_z^2 - k^2(1 - \omega_p^2/\omega^2)\right] -$$
$$\frac{\omega_p^2 \omega_e^2 (k_z^2 + k^2)}{2\omega^2 \Delta\omega^2} \pm \qquad (19)$$
$$\frac{\omega_p^2 \omega_e \sqrt{4 k_z^2 k^2 \Delta\omega^2 + \omega_e^2 (k_z^2 + k^2)^2}}{2\omega^2 |\Delta\omega^2|},$$

Where $\Delta\omega^2 = \omega^2 - \omega_e^2 - \omega_p^2$, - $\omega_e$ electronic cyclotron frequency, - an $\omega_p$ electronic plasma frequency.

As it is seen the expressions (19) do not contain a denominator $(\omega - \omega_e)$. The other expressions of this dispersion equation do not contain this denominator too. Thus, its solutions have no features, when $\omega \approx \omega_e$. As it is seen from (19) $k_1^2$ and $k_2^2$ have singularity in a point of upper hybrid frequency $\omega_h^2 = \omega_e^2 + \omega_p^2$. Squares of these wave numbers can be positive, negative and complex. The solutions of the equation (18) and their properties essentially depend on the signs $k_1^2$ and $k_2^2$.

First of all, we are interested in the dispersion in the field of frequencies close to the upper hybrid frequency $\omega_h$, when $\omega_h |\Delta\omega^2| \ll \omega^2, \omega_e^2$. In this case the expressions for $k_{1,2}^2$ become essentially simpler and have the following way:

At $\Delta\omega^2 > 0$:

$$k_1^2 = k^2 - k_z^2 - \frac{\omega_p^2}{\omega^2} \frac{k^4}{k_z^2 + k^2},$$
$$k_2^2 = k^2 - k_z^2 + \frac{\omega_p^2}{\omega^2} k^2 \left(1 + \frac{k_z^2}{k_z^2 + k^2}\right) - \qquad (20)$$
$$- \frac{\omega_p^2}{\omega^2} \frac{\omega_e^2}{\Delta\omega^2}(k_z^2 + k^2),$$

At $\Delta\omega^2 < 0$:

$$k_1^2 = k^2 - k_z^2 + \frac{\omega_p^2}{\omega^2} k^2 \left(1 + \frac{k_z^2}{k_z^2 + k^2}\right) +$$
$$+ \frac{\omega_p^2}{\omega^2} \frac{\omega_e^2}{|\Delta\omega^2|}(k_z^2 + k^2), \qquad (21)$$
$$k_2^2 = k^2 - k_z^2 - \frac{\omega_p^2}{\omega^2} \frac{k^4}{k_z^2 + k^2}.$$

From the expressions (20) – (21) it follows that near to the upper hybrid frequency the square of one of the wave numbers describing traversal structure of the field in plasma does not depend on the value of the enclosed magnetic field and when $\omega_p^2 \ll \omega^2$ (plasma of small density) is close to the square of vacuum traversal number $\kappa^2$. The second wave number in each of the fields has the strong dependence on the magnetic field, and at $|\Delta\omega^2| \to 0$ tends to infinity. At $\Delta\omega^2 > 0$, $k_2^2 \to -\infty$, and at $\Delta\omega^2 < 0$, $k_1^2 \to \infty$. In the field of frequencies, which is interesting for us, there are two small parameters: $\frac{\omega_p^2}{\omega^2} \ll 1$ and $\frac{\Delta\omega^2}{\omega_e^2} \ll 1$. From (20) – (21) it follows that, the behavior of the "plasma" traversal wave numbers depending on magnetic field will be essentially various, depending on the relation between these quantities. In the region of frequencies, where the quadratic detuning from the upper hybrid frequency is small ($|\Delta\omega^2| < \omega_p^2$) the plasma influence becomes essential. Otherwise its influence on dispersion becomes weak. In the field of frequencies $\omega > \sqrt{\omega_h^2 + \omega_p^2}$ and $\omega < \sqrt{\omega_h^2 - \omega_p^2}$ at $\omega_p^2 \ll \omega^2$ plasma wave numbers are of one order ($k_1 \sim k_2 \sim \kappa$). The structure of the field in plasma in this field of frequencies is close to the vacuum. Fig. 7 and in Fig. 8 give the dependence $k_{1,2}^2(k_z)$ for various frequencies at the following parameters: density of plasma of $n = 10^9$, $\omega_p = 1.78 \cdot 10^9$ Hz, the magnetic field of $H = 960$ gauss, $\omega_e = 1.69 \cdot 10^{10}$ Hz.

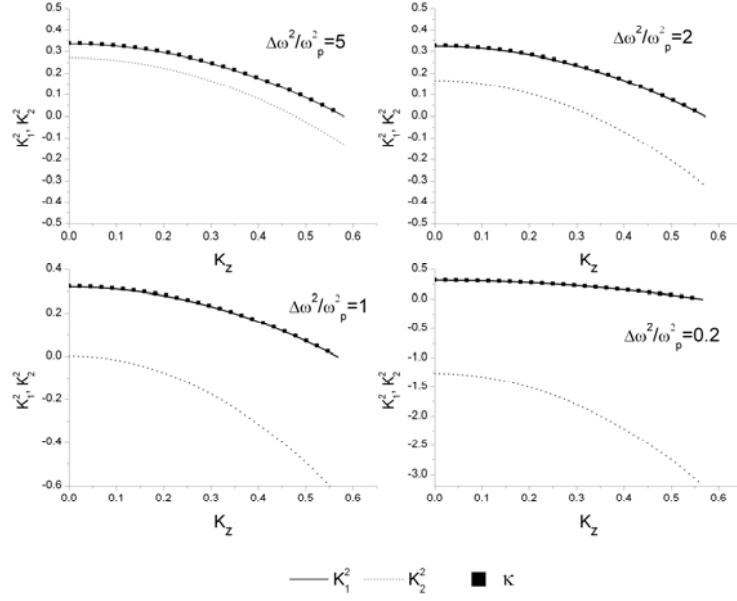

Fig. 7. Dependence of the squares of traversal wave numbers, describing the structure of the field in plasma depending on longitudinal wave number at different values of parameters $\Delta\omega^2/\omega_p^2$ for the field of frequencies above the upper hybrid.

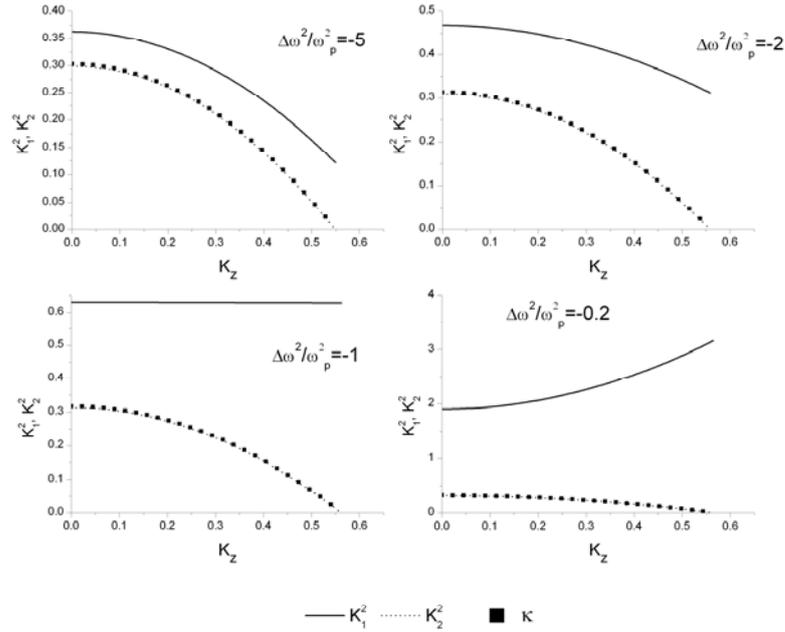

Fig. 8. Dependence of the squares of traversal wave numbers, describing the structure of the field in plasma depending on longitudinal wave number at different values of parameters $\Delta\omega^2/\omega_p^2$ for the field of frequencies below the upper hybrid one.

The dispersion equation (18) admits the possibility of analytical investigation in the range of frequencies above and below the upper hybrid frequency where the quadratic detuning does not exceed $\omega_p^2$ $|\Delta\omega^2| < \omega_p^2$ ) $\omega_p^2 \ll \omega_e^2 \sim \omega_h^2$. Let's also suppose that the following conditions are satisfied: $\kappa a > 1$, $k_i a > 1, \kappa b > 1, \; k_i b > 1$. In this case the cylindrical functions which is contained in the dispersion equation (18) can be changed by their asymptotic approaches. It has been mentioned above that in the field of the frequencies located above $\omega_h$ ( $k_1^2 \sim \kappa^2$ and $k_2^2 < 0, \; k_2^2 \to -\infty$ at $\Delta\omega^2 \to 0$), there is a finite number of solutions. In the field of the frequencies low $\omega_h$, in which the condition is satisfied $|\Delta\omega^2| < \omega_p^2 \ll \omega_e^2 \sim \omega_h^2$, and $\Delta\omega^2 < 0$, according to expressions (19) – (21) $k_2^2 \sim \kappa^2$, and

$k_1^2 > 0$, $k_1^2 \to \infty$ at $|\Delta\omega^2| \to 0$. In this case the dispersion equation (18) can be transformed into simplier form convenient for analytical and numerical investigation:

$$\sqrt{\frac{\omega_p^2}{\Delta\omega^2}\frac{k_z^2+k^2}{k^2}} ctg\left[\sqrt{\frac{\omega_p^2}{\Delta\omega^2}(k_z^2+k^2)}a - \frac{\pi}{4}\right] \times \left[\frac{J_0(\kappa a)}{J_1(\kappa a)}(k_z^2+k^2) - \left(\frac{\Delta_{01}(a)}{\Delta_{11}(a)}k_z^2 + \varepsilon_3\frac{\Delta_{00}(a)}{\Delta_{10}(a)}k^2\right)\right] =$$
$$= \frac{\kappa}{k}\left[\frac{J_0(\kappa a)}{J_1(\kappa a)}\left(\varepsilon_3\frac{\Delta_{00}(a)}{\Delta_{10}(a)}k_z^2 + \frac{\Delta_{01}(a)}{\Delta_{11}(a)}k^2\right) - \varepsilon_3\frac{\Delta_{00}(a)\Delta_{01}(a)}{\Delta_{11}(a)\Delta_{10}(a)}(k_z^2+k^2)\right],$$
(22)

where $\varepsilon_3 = 1 - \omega_p^2/\omega^2$. At every fixed $k_z$ the right part of equation (22) is slowly changing continuous function of $\Delta\omega^2$ in the range of frequencies $(-\omega_p^2, 0)$. On the other hand, the left part represents the infinite collection of the curves between asymptotes, defined by the conditions:

$$\kappa n + \pi/4 < \sqrt{\frac{\omega_p^2}{\Delta\omega^2}(k_z^2+k^2)}a < \pi(n+1) + \pi/4,$$
$$n = 0,1,2,3...$$

Thus, in the frequency band below the upper hybrid frequency ($|\Delta\omega^2| < \omega_p^2 \ll \omega_e^2 \sim \omega_h^2$) $\Delta\omega^2 < 0$ there is an infinite number of branches of natural oscillations of the wave guide partially filled with plasma. In this case these branches are placed completely in this band. Besides, there are branches which enter into this band from the frequency band $\omega < \omega_e$.

The dispersion equation (18) has been solved numerically. As it is seen from equation (22), in the narrow field of frequencies located below upper hybrid, there is an infinite number of branches situated completely in this band that is presented in Fig. 9.

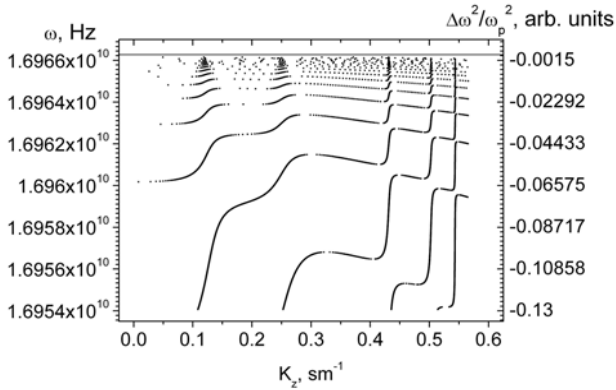

Fig. 9. Dispersion curves of the round cylindrical wave guide partially filled with plasma in narrow field of frequencies, located below the upper hybrid frequency.

Their presence is caused only by plasma influence. As it is seen from Fig. 9, the dispersion of hybrid system in this field has the complex character. It was technically inconveniently to trace separately each dispersion curve presented in Fig. 9, at the numerical solution of the dispersion equation (18). The solution of equation has been obtained relatively $k_z$ for a great number of values of parameter $\Delta\omega^2/\omega_p^2$. Corresponding points were plotted on a coordinate plane therefore the contours of dispersion curves were obtained.

## CONCLUSION

Thus, the results obtained above show a major variety of physical processes which can be observed at nonlinear interaction of waves. First of all we shall point out, that at three-wave interaction the degree of coherence of any waves can essentially increase. It essentially differs from ordinal processes when the degree of regularity of interacting waves is maintained or essentially decreases at transition in a regime with chaotic dynamics. If there is decay of HF wave on HF and LF one (the process which has been discussed in the first section) during the restricted interval of time and when the observation is conducted for HF waves it can seem, that violation of the second principle of thermodynamics takes place. The considered process can have not only general physics meaning but it can be also used for increasing of the level of radiation coherence. Certainly, plasma is not that nonlinear medium in which such processes should be realized in practice. For this purpose it is possible to use other nonlinear mediums, for example crystals.

Three-wave interaction is a simplified model of the real nonlinear interaction of waves which is much more complex. The first complication which it is necessary to pay attention to is related with the fact that dispersion of real electrodynamic systems is such that the chains of three-wave decays can take place in them. The regular dynamics of such infinite stages is investigated. In papers ([7-9]) it has been shown, that their dynamics is regular. However, electrodynamic structures allow to realize only the finite number of stages. Dynamics of the finite number of such stages has been analyzed above. Its interesting feature is that on intervals of time smaller than the time necessary for excitation of the extreme satellite the dynamics of these stages is quite regular. It coincides with the theoretical conclusions obtained for the infinite stages. However, after this time dynamics of these interactions becomes complicated essentially. It becomes complicated both in symmetrical case and especially in asymmetrical case (when the number of the blue and red satellites is various). And in that case, when requirements of synchronism are not exactly performed ($\delta \neq 0$), dynamics of such interactions, within the limits of the used model becomes chaotic.

The processes of three-wave interaction can be important not only for the understanding of physical processes but they can be used for various applications, for example for plasma heating, and also for formation of the necessary shape of spectrum radiation. In the latter case it is exclusively important the knowledge of the arrangement of the natural branches of oscillations of electrodynamic structure in which there is an interaction of waves. The more the distance (of frequencies) between the waves taking place in interaction, the greater should be the amplitudes of waves for transition in a regime with dynamic chaos. For example, as our preliminary estimations show, the intensities of the fields exceeding 60 KV/sm are necessary for radiation in a ten-cm gamut in the unbounded plasmas ( $\omega_p \sim 10^{11}\ sm^{-3}$ ) and for transition in a mode with dynamic chaos. Therefore for formation of spectrums of radiation with smaller intensities it is necessary to use the restricted plasma structures or the structures with smaller density of plasma. Such structure also has been also investigated above. It is necessary to notice, that decreasing of plasma density allows not only to reduce the necessary for stochastization field intensity, but also inevitable losses appearing from excitation of LF waves (plasma waves).